\documentclass{article}

\usepackage{arxiv}

\usepackage[utf8]{inputenc} 
\usepackage[T1]{fontenc}    
\usepackage{hyperref}       
\usepackage{url}            
\usepackage{booktabs}       
\usepackage{amsfonts}       
\usepackage{nicefrac}       
\usepackage{microtype}      

\usepackage{graphicx}

\usepackage{cite}

\graphicspath{{./Figs/}}

\usepackage{wrapfig}
\usepackage{tabu}
\usepackage{amsmath}
\DeclareMathOperator*{\argmin}{arg\,min}

\usepackage[dvipsnames]{xcolor}

\usepackage{amsthm}
\newtheorem*{definition}{Definition}

\usepackage{expl3}
\ExplSyntaxOn
\newcommand\latinabbrev[1]{
  \peek_meaning:NTF . {
    #1\@}%
  { \peek_catcode:NTF a {
      #1.\@ }%
    {#1.\@}}}
\ExplSyntaxOff
\def\eg{\latinabbrev{e.g}}
\def\etal{\latinabbrev{et al}}

\usepackage[normalem]{ulem}

\title{RGB Point Cloud Manipulation with Triangular Structures for Artistic Image Recoloring}
\author{
  Baptiste Delos \\
  IRIT, Université de Toulouse, CNRS, \\
  INPT, UT1, UT2J, UT3\\
  France\\
   \And
  Nicolas Mellado \\
  IRIT, Université de Toulouse, CNRS, \\
  INPT, UT1, UT2J, UT3\\
  France\\
  \texttt{nicolas.mellado@irit.fr} \\
   \And
  David Vanderhaeghe \\
  IRIT, Université de Toulouse, CNRS, \\
  INPT, UT1, UT2J, UT3\\
  France\\
  \texttt{vdh@irit.fr} \\
   \And
  Rémi Cozot \\
  LISIC, Université du Littoral Côte d'Opale,  \\
  France\\
  \texttt{remi.cozot@univ-littoral.fr} \\
}

\newcommand{\numStruct}{\ensuremath{k}}

\begin{document}
\maketitle
\begin{abstract}
Usual approaches for image recoloring, such as local filtering by transfer functions and global histogram remapping, lack of accurate control or miss small groups of important pixels.
In this paper, we introduce a triangle-based structuring of the colors of an image in the RGB space.
We present an analysis of image colors in the RGB space showing the theoretical motivation of our triangular abstraction. We illustrate the usefulness of our structure to recolor images.

\end{abstract}


\section{Introduction}
In the last decades, creating and manipulating pictures has become a daily activity for many people.
Professional photographers build a technical setup to obtain a desired photographic style (e.g. lighting and camera settings), and edit colors in post-process using imaging software.
Average users capture a picture and, on the fly, apply blackbox filters in real-time (e.g. smartphone applications).
In both cases, filters are designed to offer control to the users, and let them give their own style to the pictures.

\begin{figure*}[htb]
\centering
\includegraphics[width=.98\linewidth]{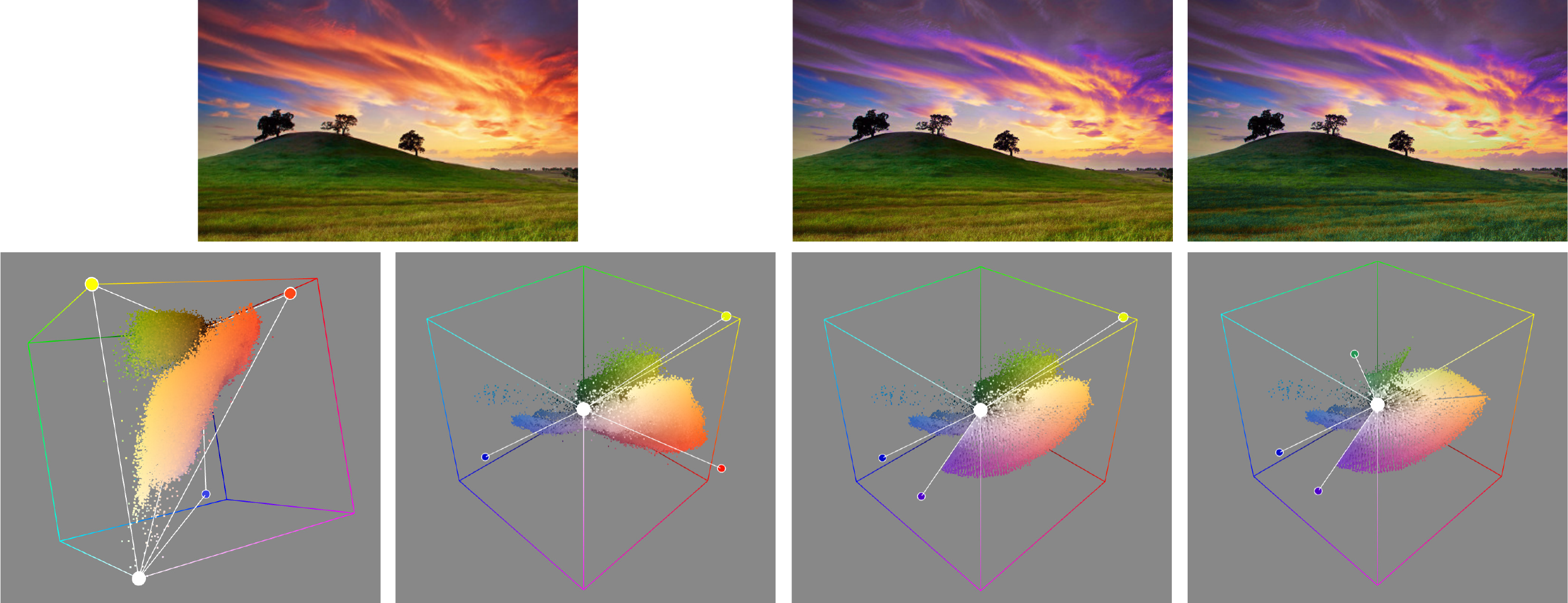}
 \centering
  \caption{\label{fig:teaser}%
  In this work, we propose to structure the colors of an image using triangles in RGB space and provide a high-level control structure for image manipulation.
  Left: an input image and its color point cloud seen from two different view points.
  The structure is composed here by three triangles sharing an edge along the illuminant axis (white to black).
  Center and right: the user modifies the structure by moving the triangles vertices, and the image colors are updated by interpolating the transformation.
  }
\end{figure*}

\begin{figure*}[htb]
\centering%
\includegraphics[width=.235\linewidth]{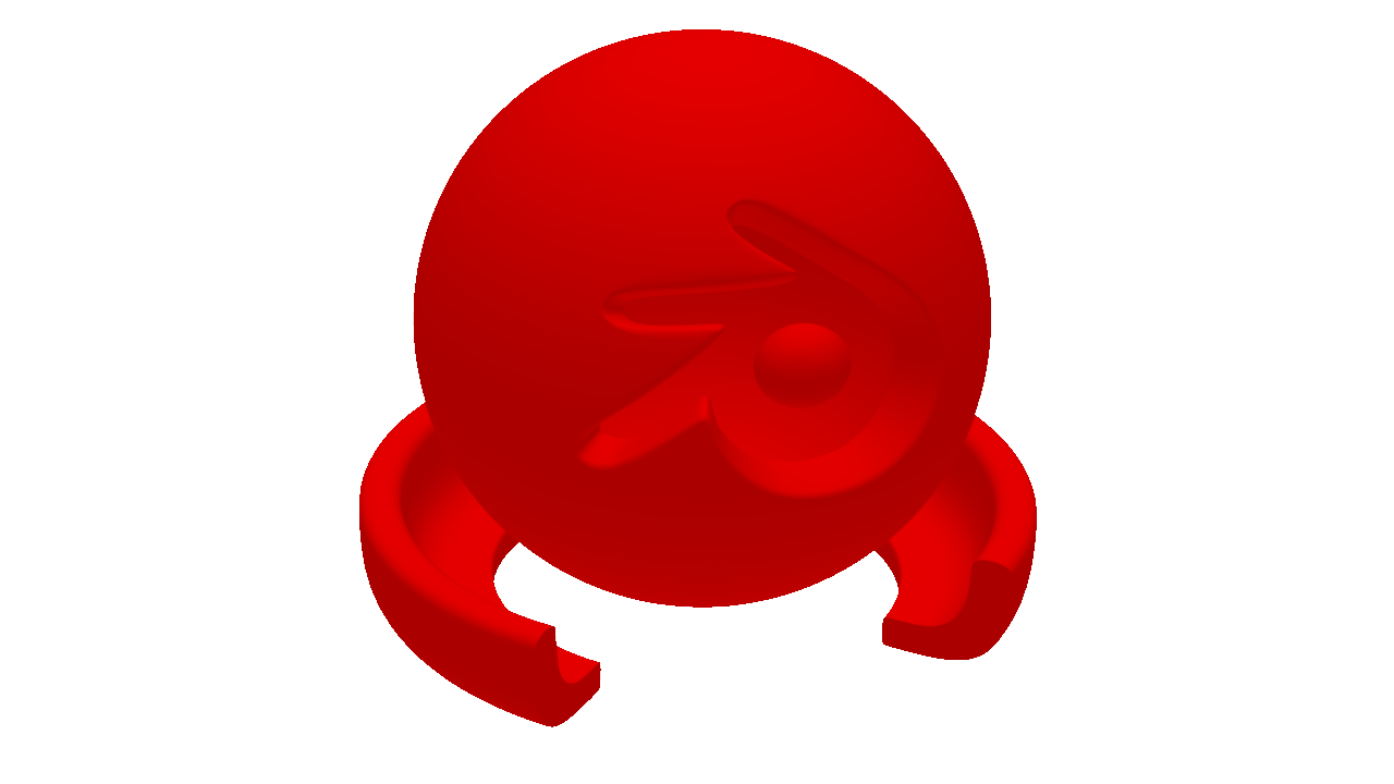}
\includegraphics[width=.235\linewidth]{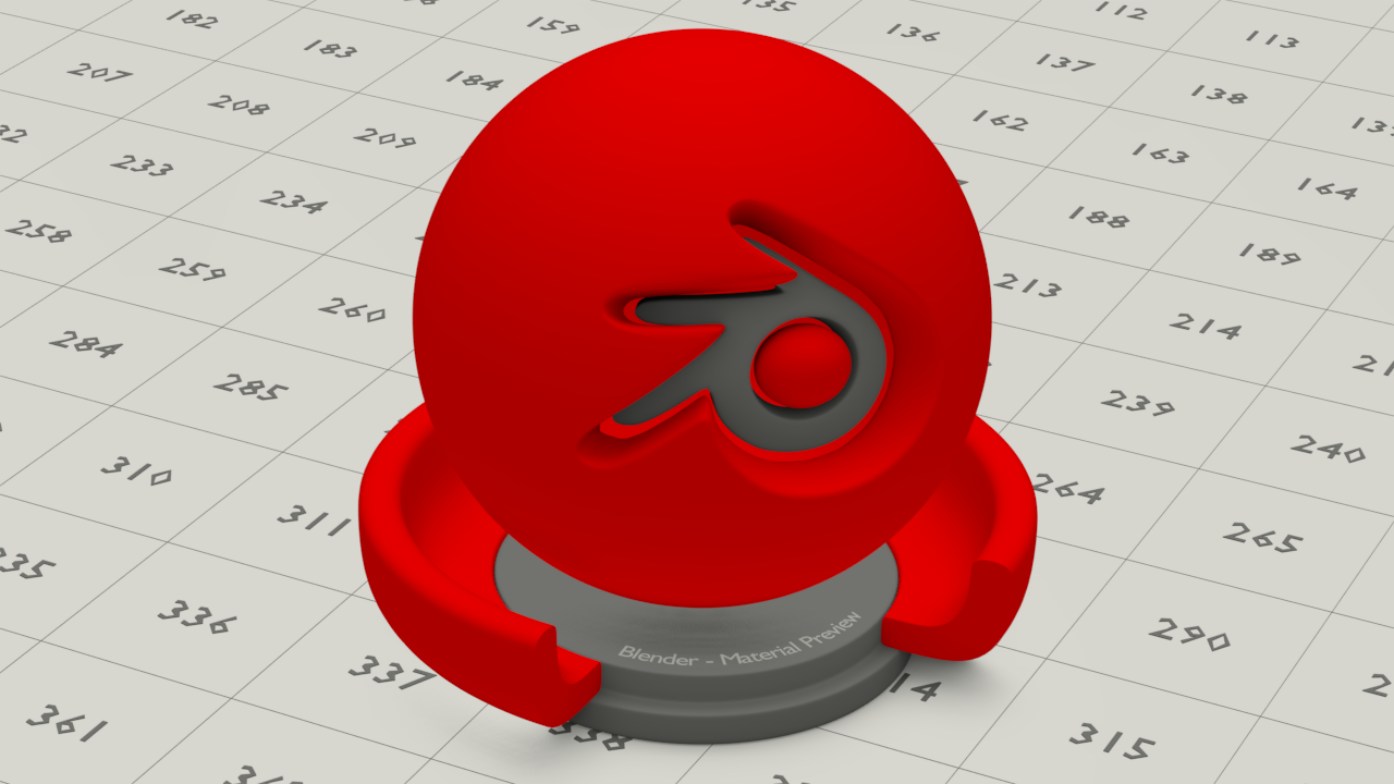}
\includegraphics[width=.235\linewidth]{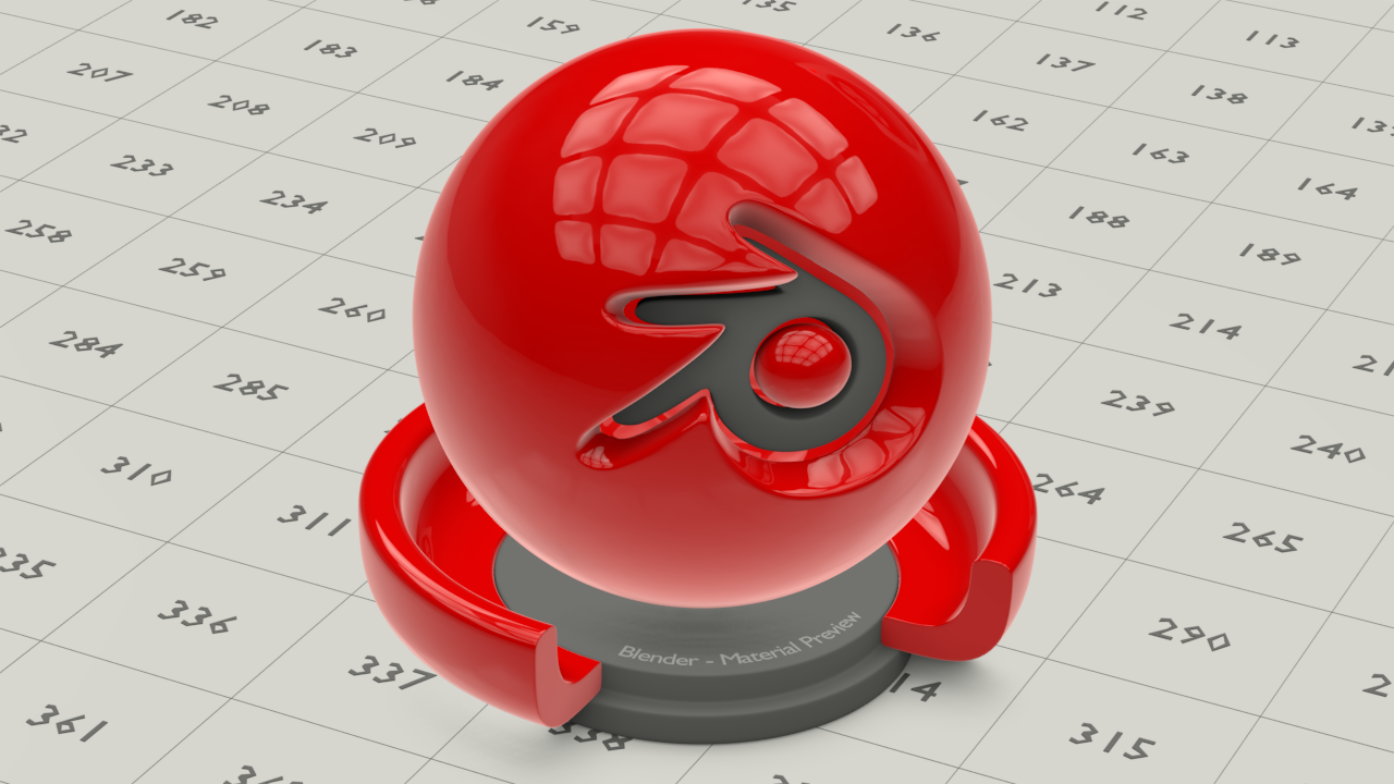}
\includegraphics[width=.235\linewidth]{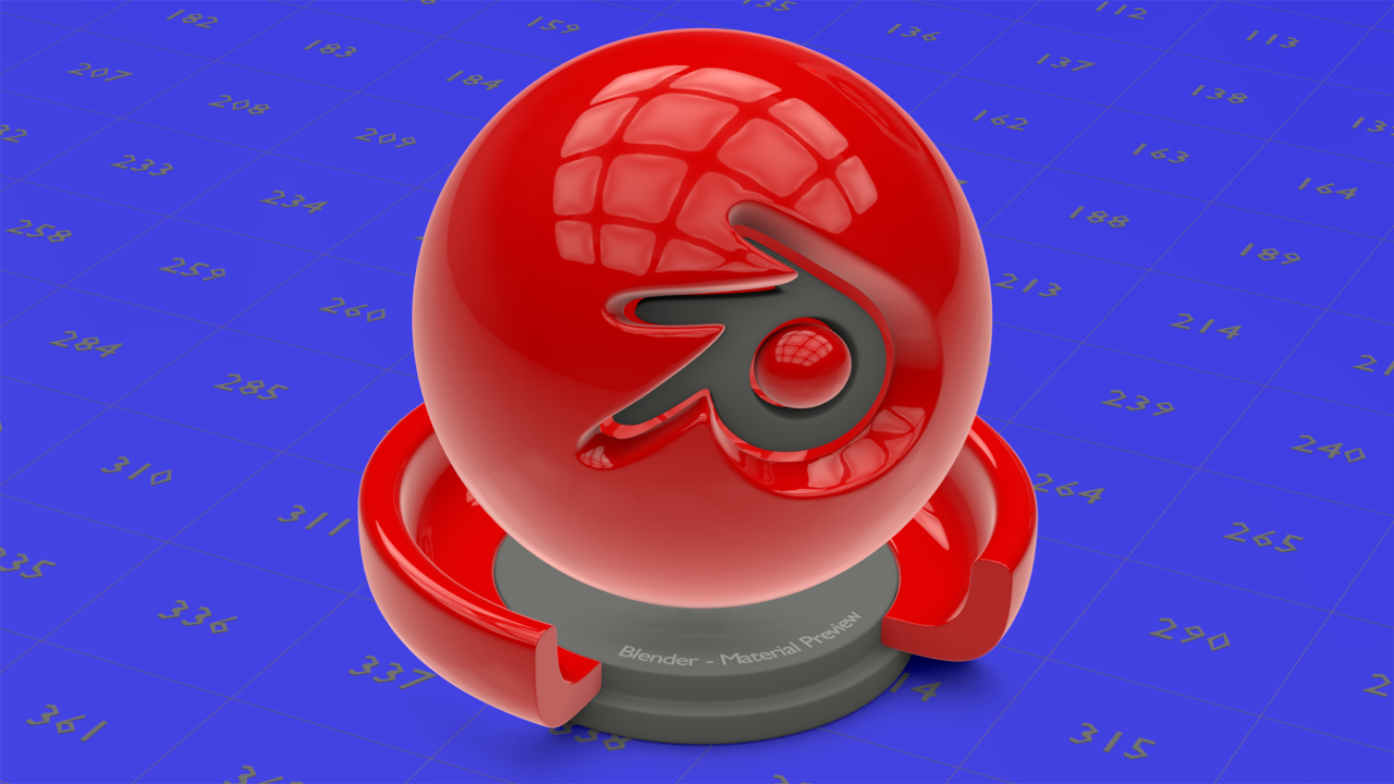}
\\
\centering%
\includegraphics[width=.235\linewidth]{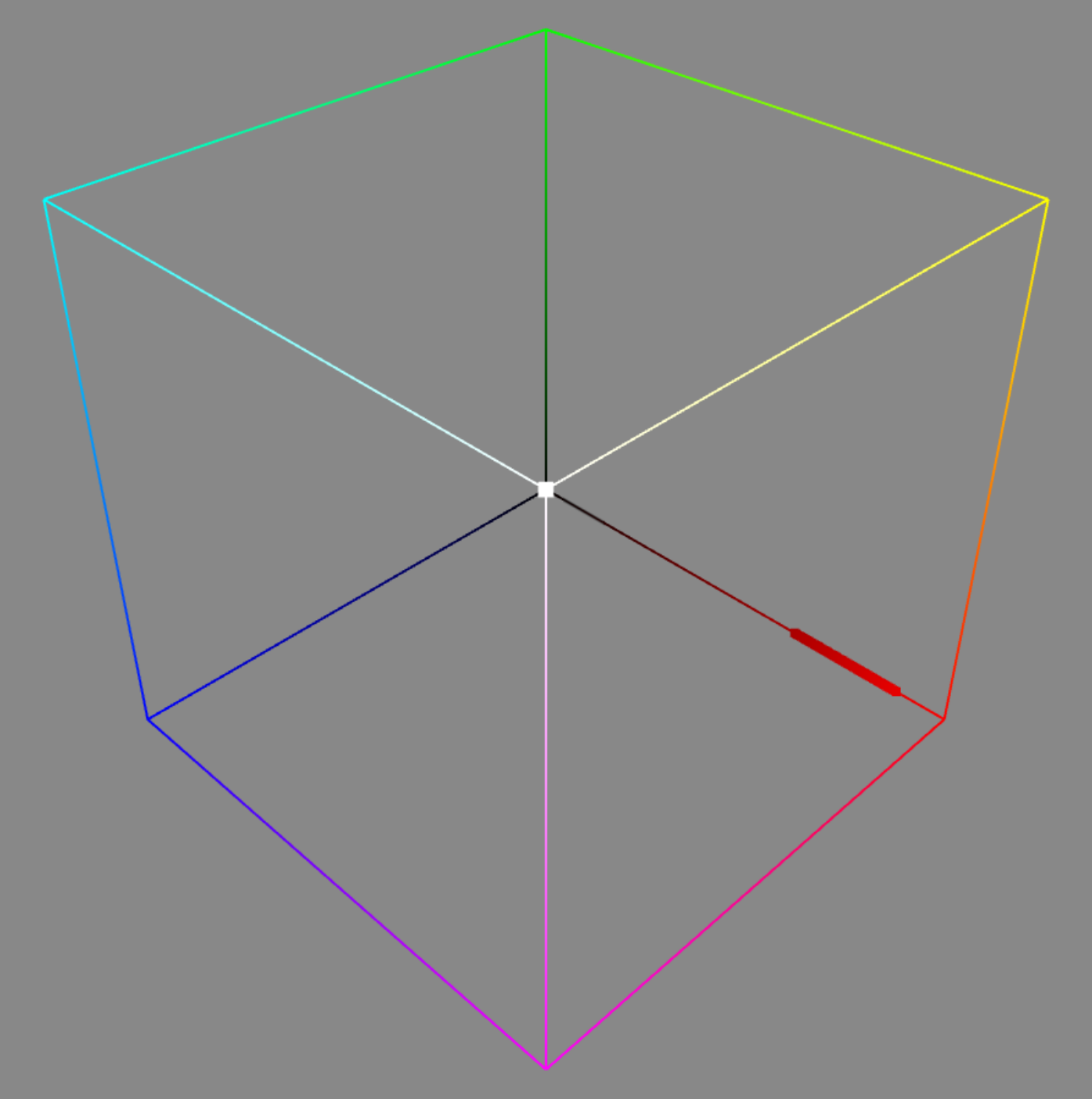}
\includegraphics[width=.235\linewidth]{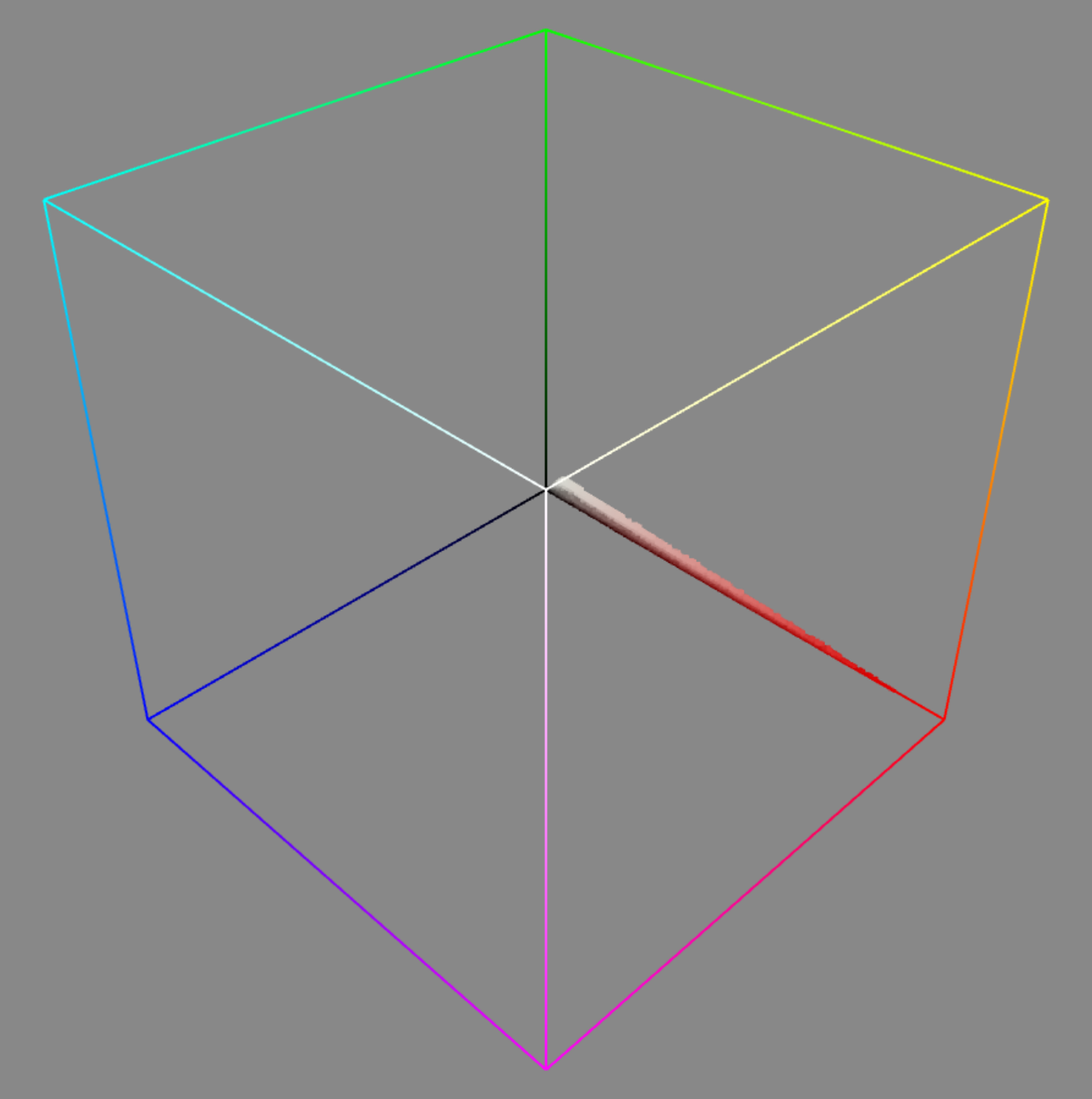}
\includegraphics[width=.235\linewidth]{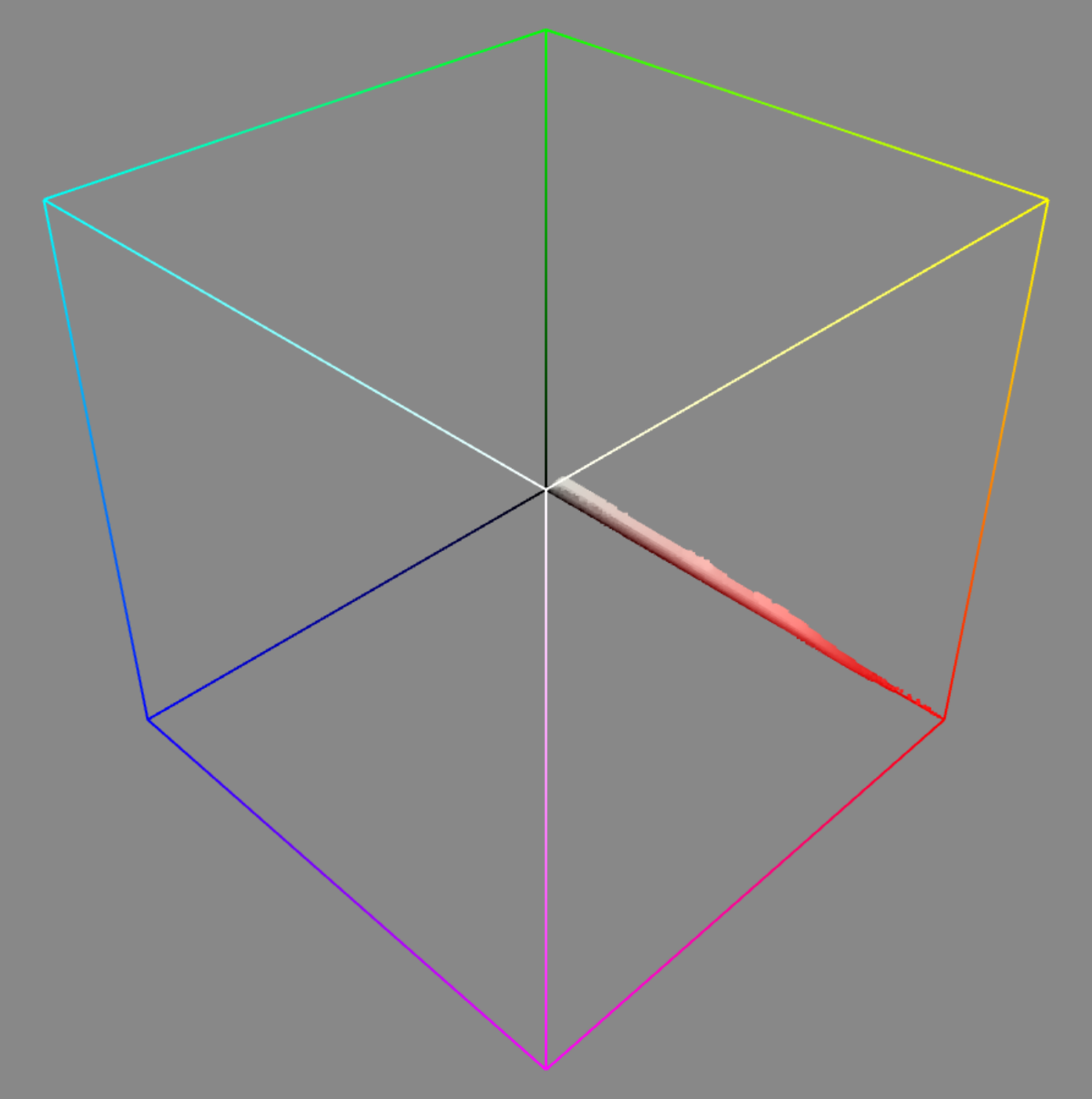}
\includegraphics[width=.235\linewidth]{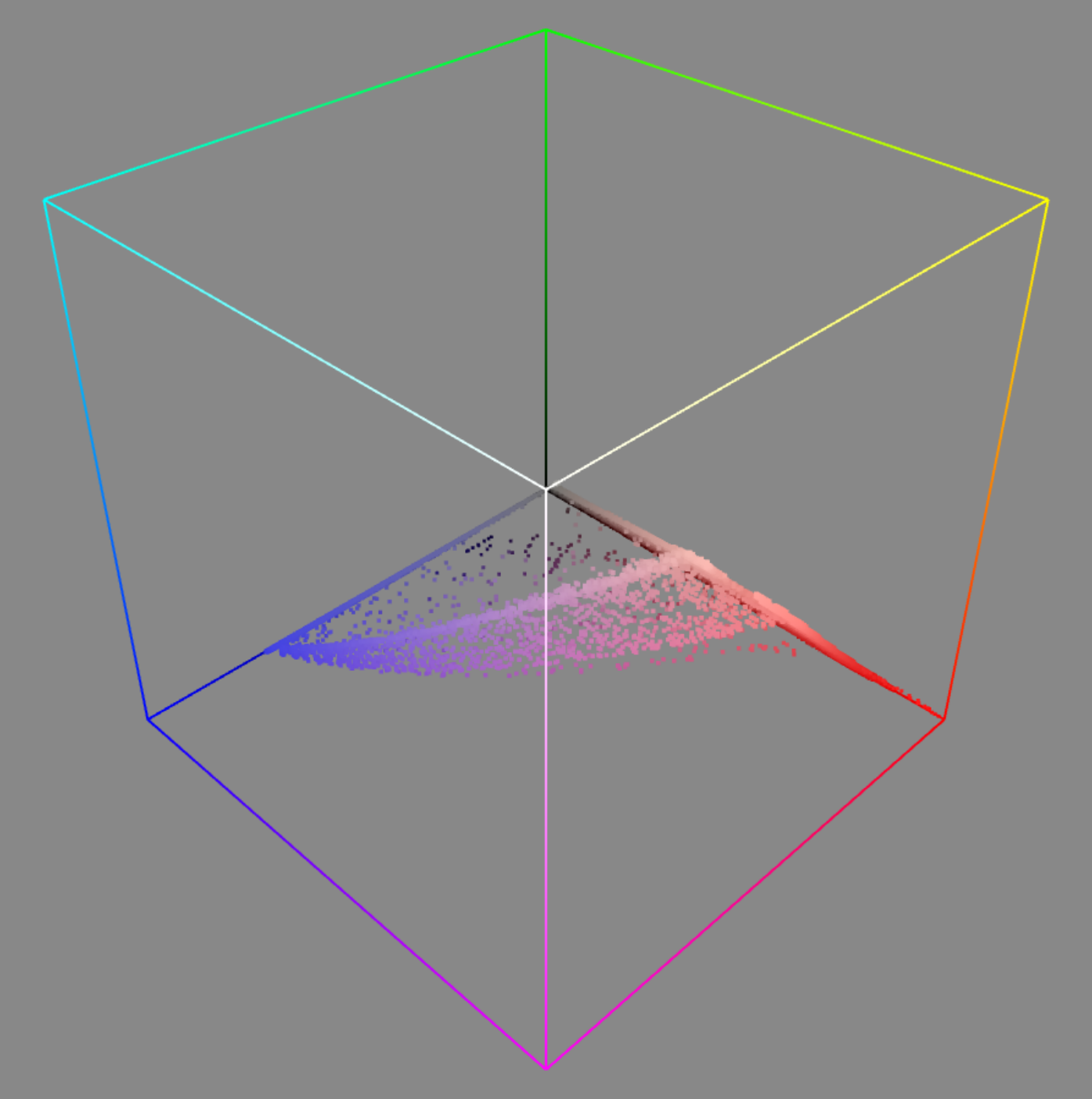}
\\
\centering%
\includegraphics[width=.235\linewidth]{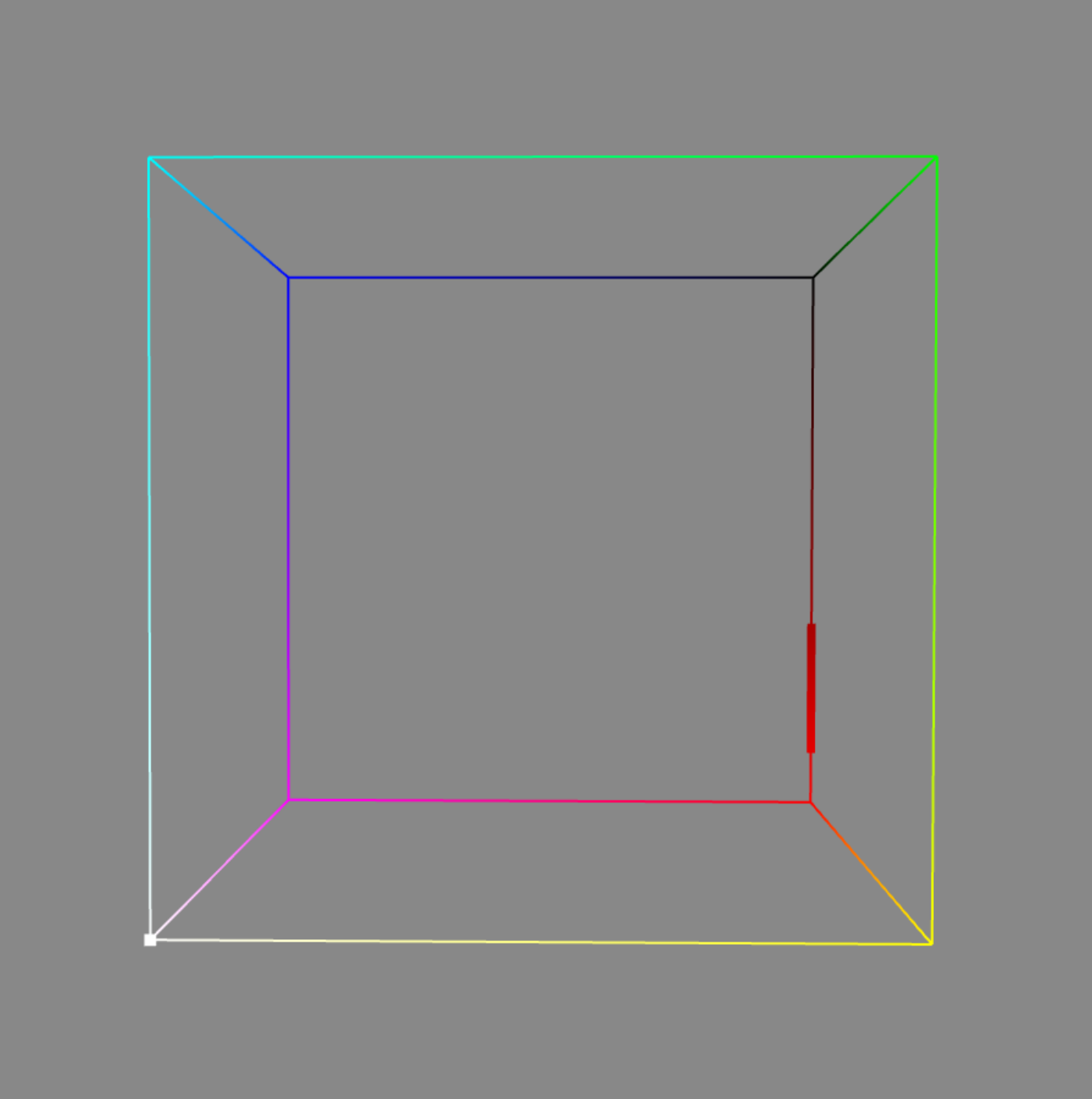}
\includegraphics[width=.235\linewidth]{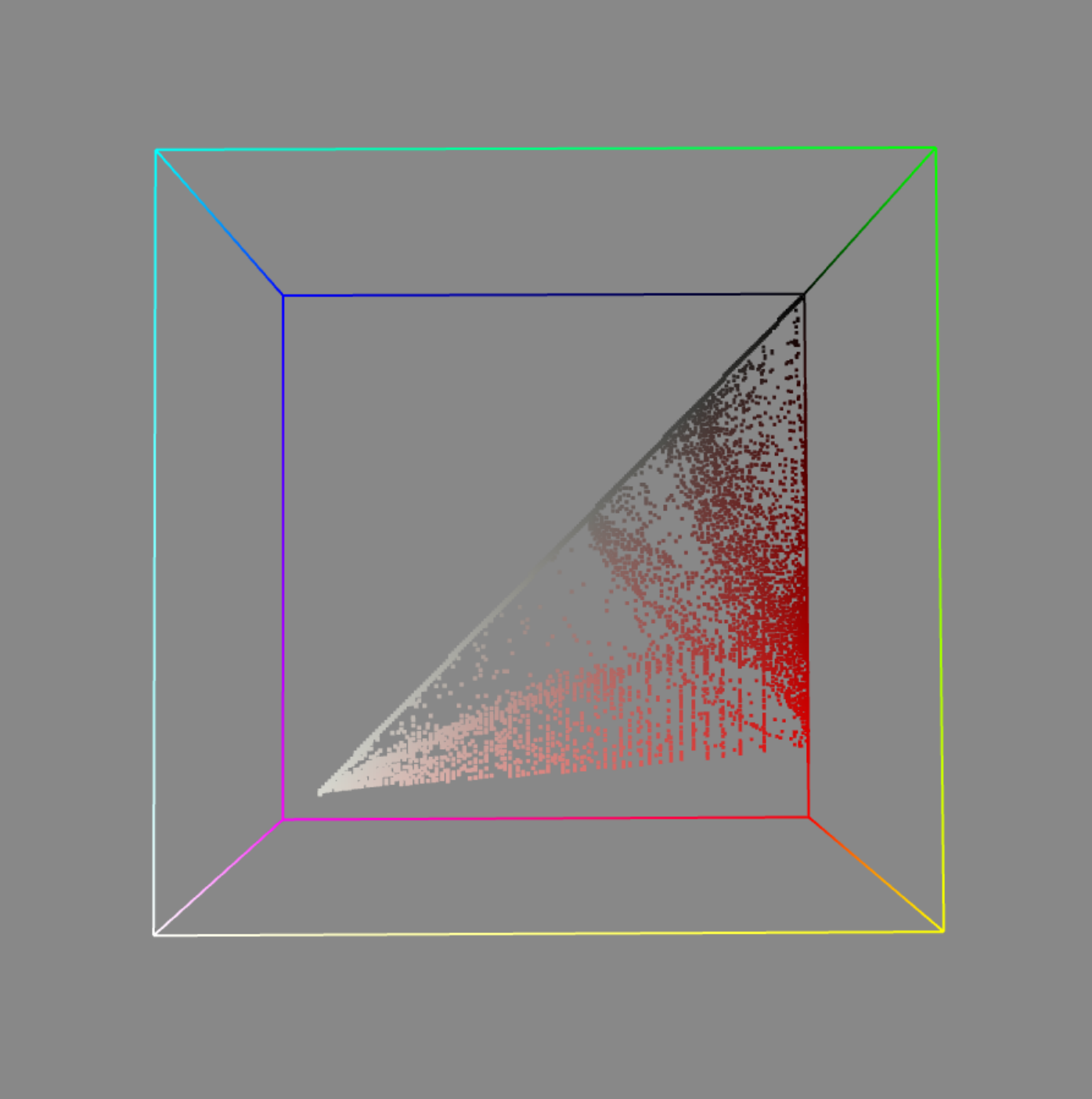}
\includegraphics[width=.235\linewidth]{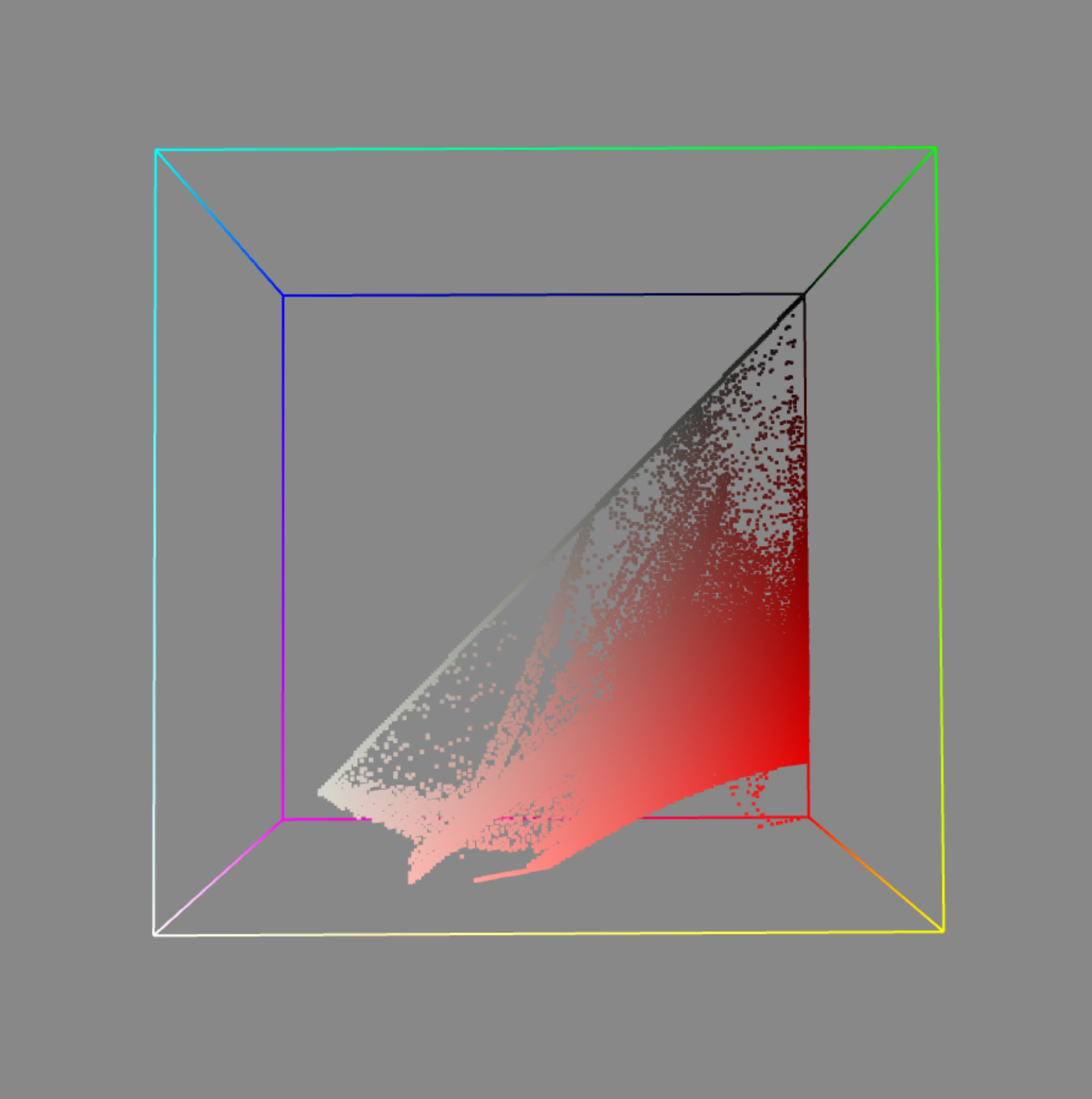}
\includegraphics[width=.235\linewidth]{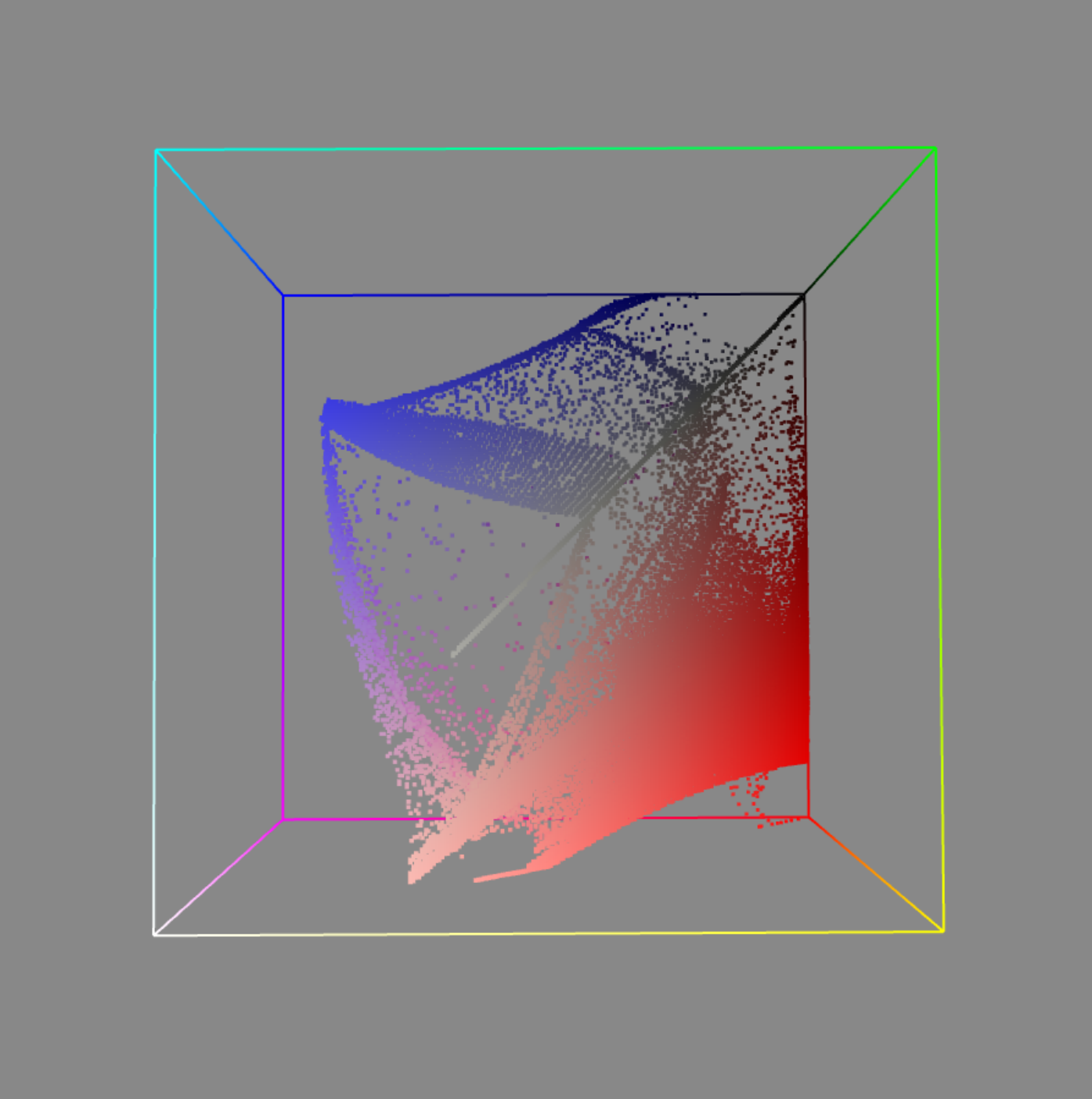}
\\
\begin{tabu} to 0.95\textwidth { X[c] X[c] X[c] X[c] }
 a) & b) & c) & d)
\end{tabu}
  \caption{\label{fig:cg}%
    Impact of the complexity of the scene on image RGB color point cloud.
    The first row shows input images, the second and third row show the same point cloud from two different viewpoints. (a) a single diffuse material, (b) a diffuse material surrounded by achromatic diffuse materials, (c) a dielectric material with a complex lighting environment, and (d) a dielectric material surrounded by diffuse materials with different color.
    }
\end{figure*}

\emph{Point operators}~\cite{Szeliski:2010:CVA:chapter3} are ubiquitous filters for image processing and computer vision.
These type of filters modify the colors of an image by considering only the pixel color, not its localization in the image nor its neighborhood.
%
The advantage of point operators lies in its computational simplicity.
Examples of such filters are color balance, exposure modification, or histogram transformations.
However, by definition points operators are not related to the image content, which makes them difficult to use for non-expert users, for instance to process local features.
To lift this limitation, users segment the image to define masks and layers, and process separately the different subsets of the image.

%

In this work, we propose to structure the colors of an image according to their distribution in color space, and to use this structure to adapt \emph{point operators} to the image content.
We observed that the colors taken from photographic images define structured point-clouds in colors spaces.
We show in Figure~\ref{fig:teaser} a picture and the point-cloud
where each point is a pixel and its coordinates are the pixel color components expressed in red, green, blue color space (RGB).
In this 3D space, different images or type of images generate different geometric patterns in the color point cloud.

Our main contribution is the definition of a structure that allows high-level control of the RGB point-cloud, where the colors are expressed and modified relatively to the structure elements.
Motivated by the rendering equation, we model the structure as a set of triangles that represents the distribution of the colors in linear RGB space (Section~\ref{sec:structure} and \ref{sec:instanciation}).
We study how to manipulate and stylize the colors of an image using this structure.
To this end we propose two types of edits (Section~\ref{sec:editing}):
(i) \emph{recoloring}: modifies the structure and update the color points accordingly, and (ii) \emph{structural filtering}: modifies the color points relatively to the structure.
We illustrate the benefits of our approach in Section~\ref{sec:results}.

\section{Previous Work}
A typical representation for color manipulations in images is to consider the color histogram.
State of the art image processing software (\eg{} Photoshop, Gimp) usually offer color histogram transformation and simple combinations with predefined layers, we refer the reader to introductory books~\cite{Solomon:2010:FDI,Szeliski:2010:CVA} for more details on the topic.
A limitation of histogram-based color processing is to rely only on the color samples density, which makes histograms tailored for large clusters of pixel with similar colors.
Small sets of colors are difficult to track in histograms, even if they have a strong impact in the final image.
In this work, we aim at analyzing the color relations through their geometrical configurations in color space, taking not only the density but also the shape of the distribution.

Analyzing the geometry of images color point clouds has raised increasing attention in the last years.
Tan and colleagues propose to decompose color images as layers by analyzing a simplified convex-hull of the image color point-cloud~\cite{Tan:2016}.
The main limitation of this work is to consider only the convex hull of the observable structures.
As a result, the color distribution inside the convex hull is not explicitly modeled nor controllable.
In contrary, our approach characterizes the enclosed structures, which encodes additional information on the image.
Duchêne~\etal~\cite{Duchene:2017:MIA} analyze the color point cloud of images in CIELab and extract the illuminant information in order to perform color grading.
Their approach allows efficient color grading without prior knowledge but is limited to illuminant edits while we provide more versatile tools.

A color palette can also be seen as the high-level structure of an image point cloud.
Several approaches have been proposed to edit photos using color palettes~\cite{Chang:2015:PPR,Mellado:2017:CPE,Tan:2018:EPD}.
In these approaches, the user modifies the palette colors, and the colors of the image are updated accordingly.
In other word, the color palette is used as a control structure for image recoloring, each of the color of the palette act as a control vertex of the RGB space.
From our observations, color palettes are however too simple and cannot represent the geometry of complex and realistic color point clouds.
In contrast, we introduce a new triangular structure that allows transformation not only on triangle vertices, but also to control the distribution of color regarding triangle surface.


\section{Method}
We denote $\mathcal{C} = \left\{\mathbf{c}_0 \ldots \mathbf{c}_n\right\}$ the color point cloud of an image, with $\mathbf{c}_i$ the coordinates of the $i^\text{th}$ image pixel in RGB space.
When acquired using a digital camera or computed using light simulation algorithms, the colors of an image represent the response of observed materials for a given lighting environment.
We propose a new structuring of RGB space as a finite set of geometric primitives.
A primitive represents the typical response of BRDF to lighting (see Section~\ref{sec:structure}).
Despite its apparent simplicity, we demonstrate how the proposed structure can also represents real materials response using a distributions around the primitives~\ref{sec:instanciation}.
Then, we show how acquired or rendered images can be edited using the proposed structure with simple yet powerful editing tools~\ref{sec:editing}.

\subsection{Triangular Structure Model}
\label{sec:structure}
As stated by the rendering equation~\cite{Kajiya:1986:RE}, the visible color of an object depends on the incoming radiance, the material response, and the geometric configuration of the scene.
The rendering equation is defined as
\begin{equation*}
L_o(\mathbf{x}, v)  = \int_{\Omega} L_i(\mathbf{x}, l) \; \rho(l, v) \; n . l \;\mathrm{d} l ,
\end{equation*}
\begin{wrapfigure}{r}{0.2\linewidth}
  \vspace{-3mm}
  \def\svgwidth{0.2\columnwidth}
\begingroup%
  \makeatletter%
  \providecommand\color[2][]{%
    \errmessage{(Inkscape) Color is used for the text in Inkscape, but the package 'color.sty' is not loaded}%
    \renewcommand\color[2][]{}%
  }%
  \providecommand\transparent[1]{%
    \errmessage{(Inkscape) Transparency is used (non-zero) for the text in Inkscape, but the package 'transparent.sty' is not loaded}%
    \renewcommand\transparent[1]{}%
  }%
  \providecommand\rotatebox[2]{#2}%
  \newcommand*\fsize{\dimexpr\f@size pt\relax}%
  \newcommand*\lineheight[1]{\fontsize{\fsize}{#1\fsize}\selectfont}%
  \ifx\svgwidth\undefined%
    \setlength{\unitlength}{132.10581622bp}%
    \ifx\svgscale\undefined%
      \relax%
    \else%
      \setlength{\unitlength}{\unitlength * \real{\svgscale}}%
    \fi%
  \else%
    \setlength{\unitlength}{\svgwidth}%
  \fi%
  \global\let\svgwidth\undefined%
  \global\let\svgscale\undefined%
  \makeatother%
  \begin{picture}(1,1.025807)%
    \lineheight{1}%
    \setlength\tabcolsep{0pt}%
    \put(0,0){\includegraphics[width=\unitlength,page=1]{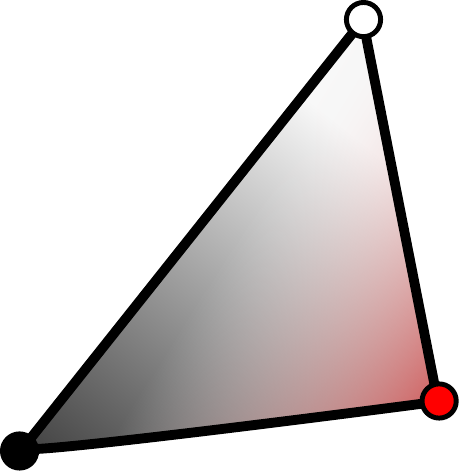}}%
    \put(0.55629486,0.26404203){\color[rgb]{0,0,0}\makebox(0,0)[t]{\lineheight{1.25}\smash{\begin{tabular}[t]{c}$\scriptstyle L_i(\mathbf{x},l) \; \rho(l,v) \; n . l$\end{tabular}}}}%
    \put(0.95204153,0.04709292){\color[rgb]{0,0,0}\makebox(0,0)[t]{\lineheight{1.25}\smash{\begin{tabular}[t]{c}$\scriptstyle \mathbf{c}_d \odot \mathbf{c}_{l}$\end{tabular}}}}%
    \put(0.72616041,0.96194731){\color[rgb]{0,0,0}\makebox(0,0)[rt]{\lineheight{1.25}\smash{\begin{tabular}[t]{r}$\scriptstyle \mathbf{c}_{s} \odot \mathbf{c}_l$\end{tabular}}}}%
  \end{picture}%
\endgroup%

  \vspace{-5mm}
\end{wrapfigure}
with $v$ and $l$ the view and light directions, $\mathbf{x}$ the observed point, $n$ the normal at $\mathbf{x}$, $\Omega$ the hemisphere around $\mathbf{x}$ supported by $n$, and $\rho$ the BRDF of the surface.
Without loss of generality, we assume the image is rendered in linear RGB
space by combining contributions for each channel (the same logic stands for XYZ space or spectral representations).
We note $\odot$ the component-wise product between two colors.
For the sake of simplicity, we assume that the BRDF
defines a diffuse color $\mathbf{c}_{d}$ and a specular color $\mathbf{c}_s$, e.g. Phong or Cook-Torrance.
We also denote $\mathbf{c}_{l}$ the color of the incoming light $L_i(x, l)$.
By construction, for uniform light and material properties, the visible colors of an object span a triangle in RGB space where vertices colors are black, $\mathbf{c}_{d} \odot \mathbf{c}_{l}$ and $\mathbf{c}_{s} \odot \mathbf{c}_{l}$, as illustrated in the right inset.

\begin{definition}{}
A Triangular Structure is a set of triangles defined in color space, and sharing an edge called the \emph{illuminant axis}.
For each triangle, the third vertex is called the \emph{colored vertex} and noted $t_i$ for the $i^\text{th}$ triangle.
\end{definition}

In real life scenarios, visible colors may deviate from the triangles, due to variations of the material and light properties, and the interactions between nearby objects (e.g. reflections, color bleeding).
Let's consider the images shown Figure~\ref{fig:cg}:
When the material is pure diffuse, without interaction with other materials, the point cloud consists in a single line between black and  $\mathbf{c}_{d} \odot \mathbf{c}_{l}$.
Adding objects in the scene creates other lines in the point cloud, with points between the lines due to antialiasing and color bleeding.
In this specific case, where a reddish material interacts with gray materials, the color bleeding fall onto the triangle defined by the reddish material.
Adding specular reflections leads to changes in the distribution of the colors within the triangle in the direction of $\mathbf{c}_{s} \odot \mathbf{c}_{l}$, but does not affect the shape of the triangle.
In the last example, the blue floor color bleeding generates interpolated colors between red and blue.

\begin{figure}[hb]
\centering
\includegraphics[height=2.8cm]{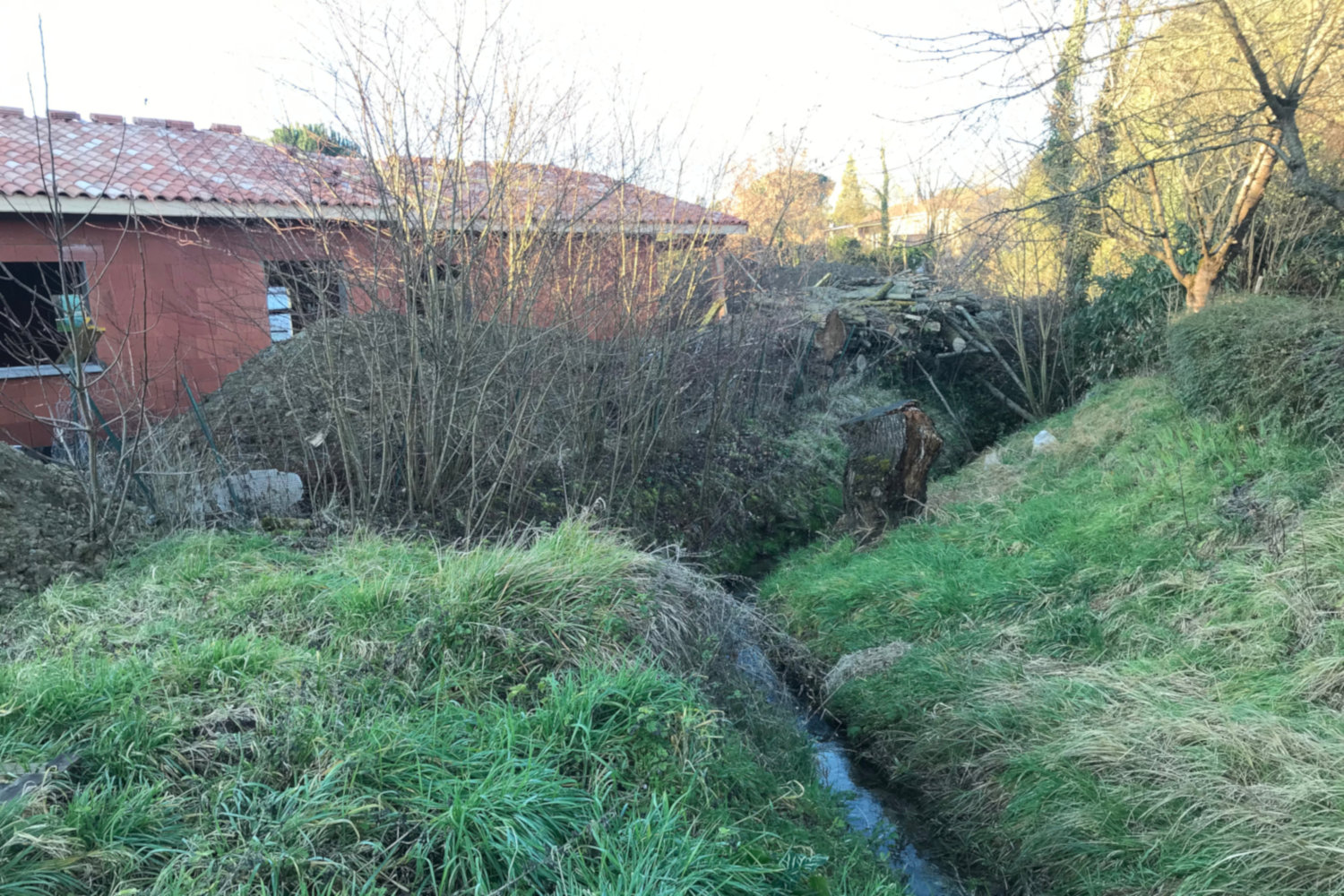}
\includegraphics[height=2.8cm]{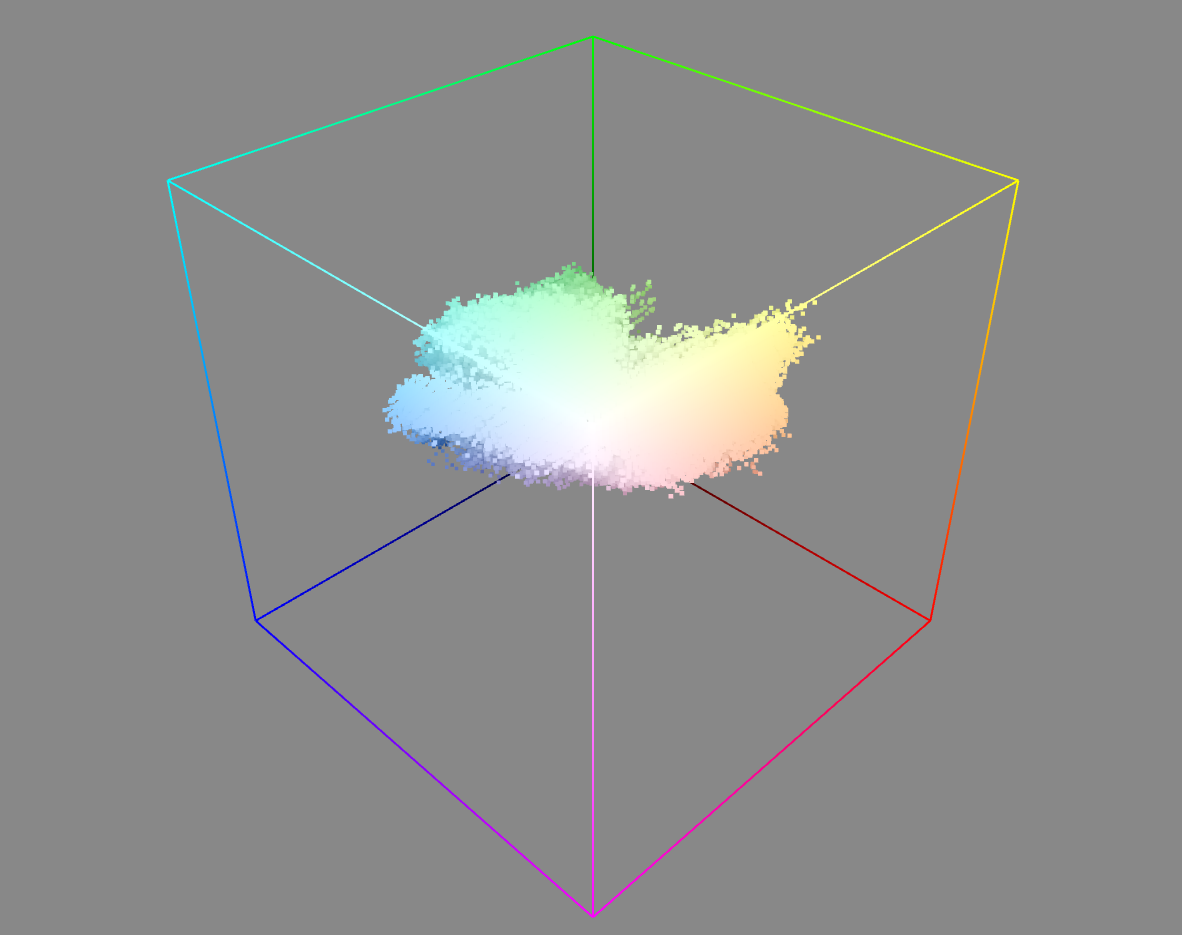}
\includegraphics[height=2.8cm]{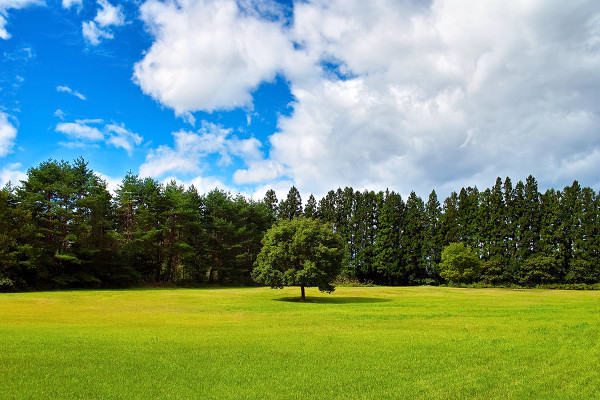}
\includegraphics[height=2.8cm]{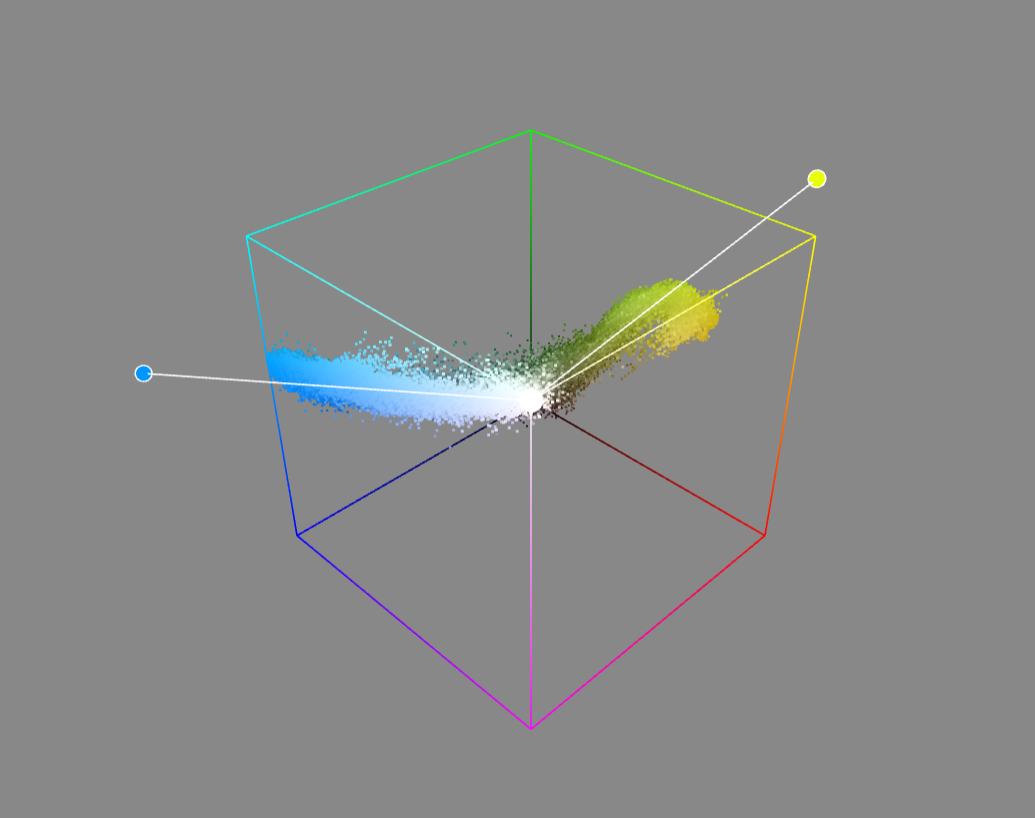}
\caption{\label{fig:colorpointscloud}%
Two pictures and the corresponding point-clouds.
The top picture, taken by an amateur, present a messy color point cloud, while the bottom picture presents more appealing colors, and more structured point cloud.
The colors associated to the most representative triangles form thin volumetric point clouds, and encode the variations of colors due to textures and lighting changes.
}
\end{figure}

More complex scenes and photographs involve stronger interactions between materials, and variations of appearance.
As shown in Figure~\ref{fig:colorpointscloud}, the variations of texture and lighting in the grass and sky areas make the colors deviate from the triangles, and form a thicker, yet clearly identifiable, volumetric point-cloud surrounding the theoretical triangles.
Other properties of the images (complex materials and lighting environment, noise, motion, drawings) might also affect the shape of the point-cloud.
Images involving multiple colored light sources can be represented by combining multiple triangular structures, however this is beyond the scope of this paper.


\subsection{Structure Setup}
\label{sec:instanciation}
In order to extract a triangular structure from a color point cloud, we propose to optimize the set of colored vertex $\mathbf{T} = \{ t_0 \ldots{} t_{\numStruct{}-1} \}$ of the structure triangles according to the color points in RGB space, by finding
\begin{equation*}
    \argmin_{\mathbf{S}, \mathbf{T}} \; \sum^{\numStruct}_{i=1} \sum_{\mathbf{c} \in S_i} d_i(\mathbf{c})^2
\end{equation*}
with $\mathbf{S} = \{ S_0 \ldots{} S_{\numStruct{}-1} \}$ the set of colors associated to each triangle, and $d_i(c)$ the distance between the color $\mathbf{c}$ and the triangle $i$.
%
%
We let the user specify the \emph{illuminant} axis, the number \numStruct{} of triangles and their initial colored vertex position.

For all our experiments we used a $k$-means approach, and optimize only the triangle orientation by rotating the colored vertex around the illuminant axis.
First, each input color point $\mathbf{c}_i$ is assigned to its nearest triangle in RGB space. 
Then, we compute for each triangle the centroid of its assigned colors and rotates its reference point so that the centroid belongs to the triangle.
The process repeats assignment/rotation steps until convergence, measured by the angular motion of the triangles at each iteration.

\subsection{Image Editing}
\label{sec:editing}

\begin{figure}[tb]
\centering
\small
\def\svgwidth{\linewidth}
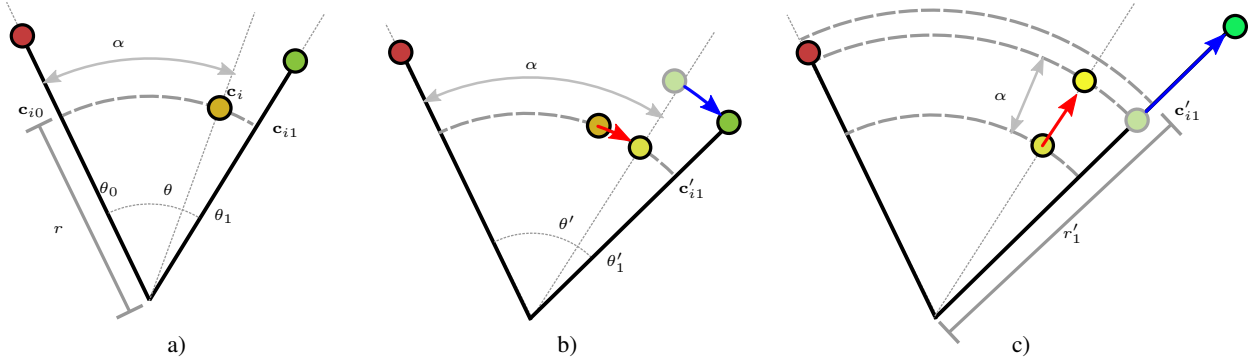
\caption{\label{fig:transform}%
A schematic 2D representation of color point transformation.
(a) Initial configuration, the yellow color point ($\theta$, $r$) is inbetween the two structures element with angle $\theta_0$, $\theta_1$.
$\alpha$ corresponds to the arc length ($\in [0, 1])$ along the arc-circle.
(b) the green colored vertex is turned clockwise (blue arrow), so the yellow point turn following the red arrow and gets greenish. It's new cylindrical coordinates $\theta'$ is  $\theta' = (1-\alpha)\theta_0+\alpha\theta'_1$.
(c) the green colored vertex is pulled away (blue arrow) the illuminant axis , the new yellow point radius is $r' = (1-\alpha)r +\alpha r'_1$.
}
\end{figure}

We define two types of transformations, namely \emph{recoloring} and \emph{structural filtering}.

\textbf{Recoloring}
We call \emph{recoloring} the process of editing the colored vertices of the triangular structure.
Let's consider a pixel color $\mathbf{c}_i$ surrounded by two triangles.
We model the transformation applied to $\mathbf{c}_i$ when moving the colored vertex of a triangle using cylindrical interpolation.
First, we express $\mathbf{c}_i$ in cylindrical coordinates $\mathbf{c}_i = (\theta, r, h)$, where $\theta$ is the angle around the illuminant axis, $r$ is the distance to the axis and $h$ the distance along the axis.
We note $\theta_0$ and $\theta_1$ the angles of the two triangles in the cylindrical domain, and $\mathbf{c}_{i0} = (\theta_0, r, h)$, $\mathbf{c}_{i1} = (\theta_1, r, h)$ the cylindrical projection of $\mathbf{c}_i$ on the triangles.
%
%
When transforming the triangles, the colors
$\mathbf{c}_{i0}$ and
$\mathbf{c}_{i1}$ are updated according to their barycentric coordinates in the triangles and noted $\mathbf{c}'_{i0}$, $\mathbf{c}'_{i1}$.
After edit, the new cylindrical coordinates are updated from the barycentric coordinates such that
$\mathbf{c}'_{i0} = (\theta'_0, r'_0, h'_0)$ and
$\mathbf{c}'_{i1} = (\theta'_1, r'_1, h'_1)$.
The final color of the considered pixel is computed by linearly interpolating the cylindrical parameters of the colors:
$\mathbf{c}'_{i} = (1-\alpha)\mathbf{c}'_{i0} + \alpha \mathbf{c}'_{i1} $, with
$\alpha = \left(\theta-\theta_0\right)/\left(\theta_1-\theta_0\right)$
(see Figure~\ref{fig:transform}).

\textbf{Structural filtering} is motivated by the observation of photographs taken by professional.
Such pictures generate point-clouds with thin elements, in contrast to ``random''  scenes that lead to thicker components (see Figure~\ref{fig:colorpointscloud}).
We design \emph{structural filtering} so the distance between a color point and its triangle is scaled by a factor defined by the user.
When the factor is zero, all color points lies on the triangle.
When the factor is greater than one, color points spread away from the triangle, covering a larger area of color space.

After transformation, color point that falls outside the gamut are reprojected on the gamut boundaries.

\section{Results}
\label{sec:results}
Direct manipulation of the vertices of the structures have different impact on the color distribution of the image.
Each of the three vertices of a structure triangle has a different \emph{meaning}.
At initialization, there is one white, one black and one color vertex.
%
The color vertices, when moved to or away the illuminant axis, controls the vividness of color set, and when rotated around the illuminant axis, it shifts the hue.
These manipulations are performed by structure element and give a control adapted to the color content of the input image.

\begin{figure}[b]
\centering%
\vspace{1em}
\includegraphics[width=.32\linewidth]{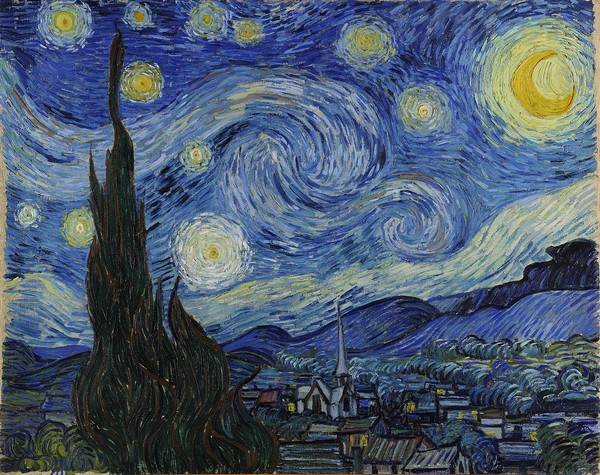}
\includegraphics[width=.32\linewidth]{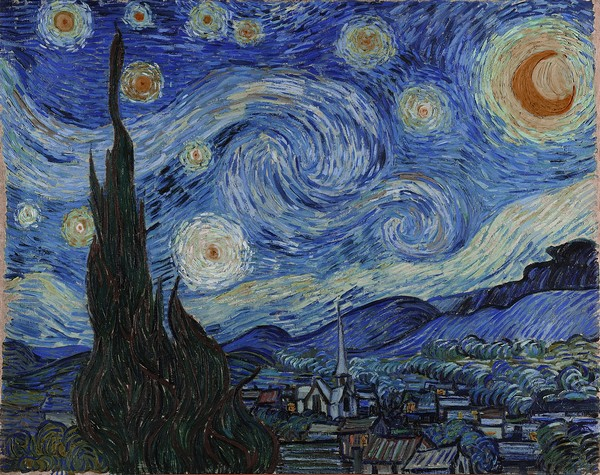}
\includegraphics[width=.32\linewidth]{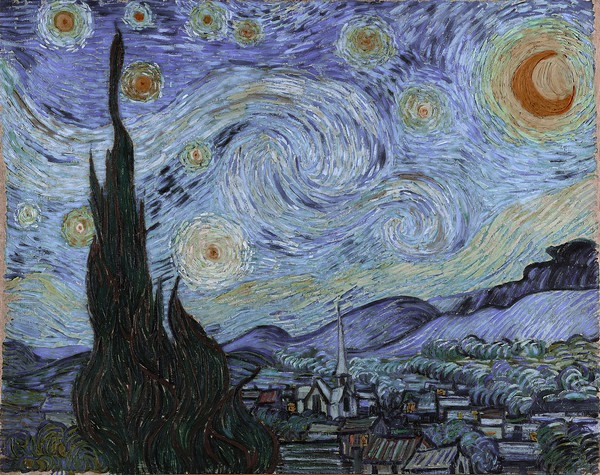}\\
\centering
\includegraphics[width=.32\linewidth]{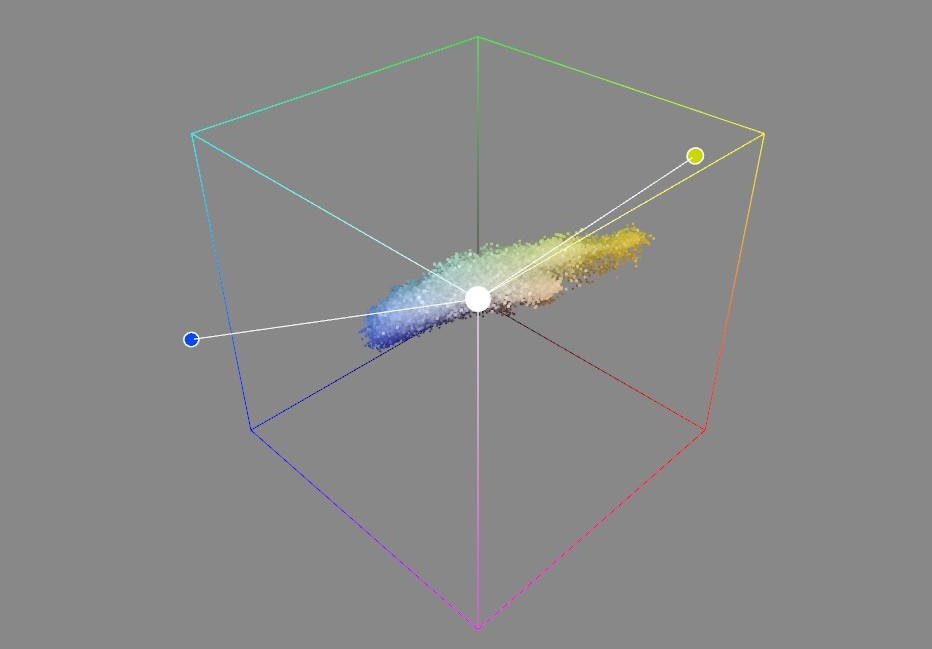}
\includegraphics[width=.32\linewidth]{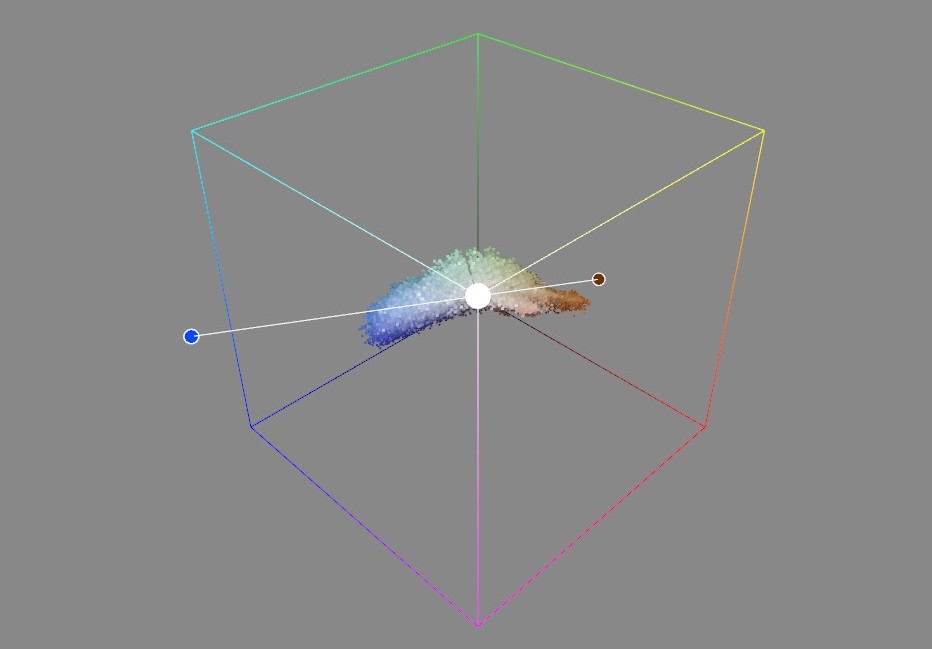}
\includegraphics[width=.32\linewidth]{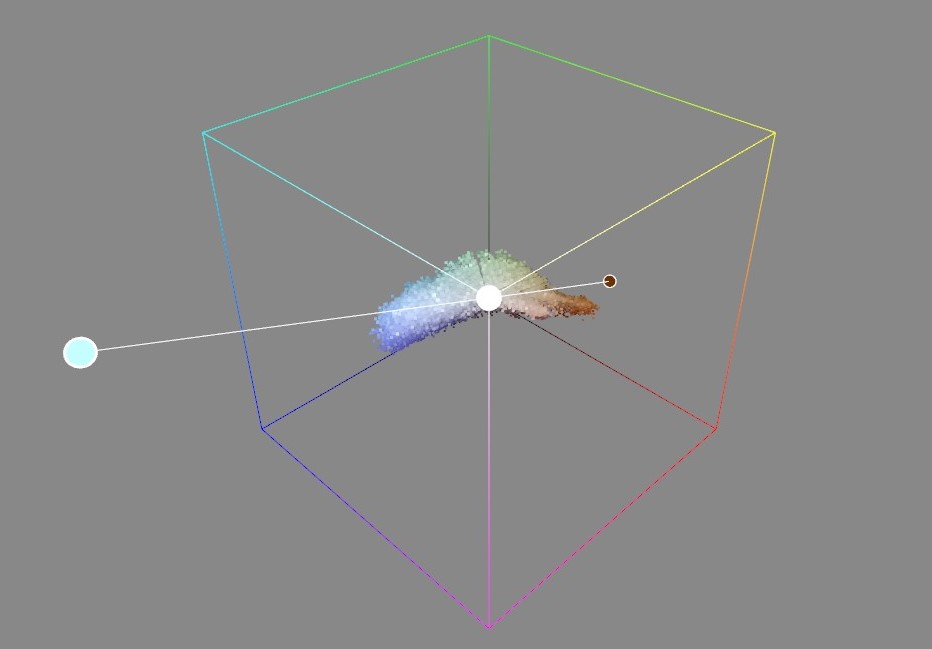}
\caption{
\label{fig:gogh} Left: input image. Center: reduction of saturation/lightness for the component yellow. Right: increased saturation for the component blue.}
\end{figure}

\textbf{Changing distances between two colored vertices}:
Figure~\ref{fig:gogh} shows edits where colored vertices of two triangles are moved along within the support plane of the structure triangle.
It controls the vividness of the colors.

\begin{figure}[t]
\centering%
 \includegraphics[width=.3\linewidth]{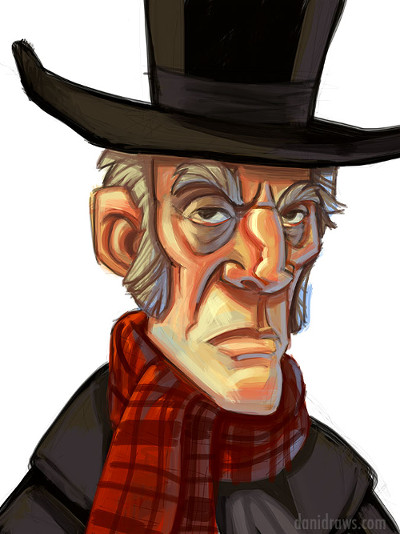}
 \includegraphics[width=.3\linewidth]{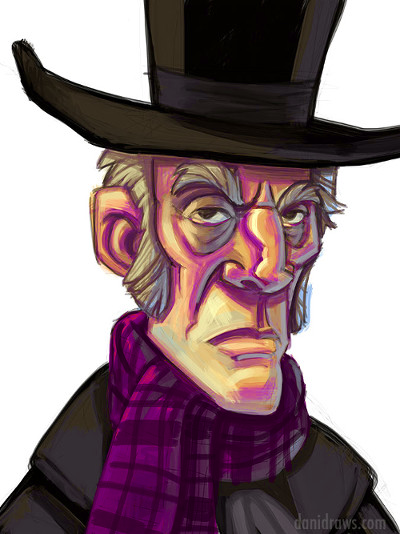}
 \includegraphics[width=.3\linewidth]{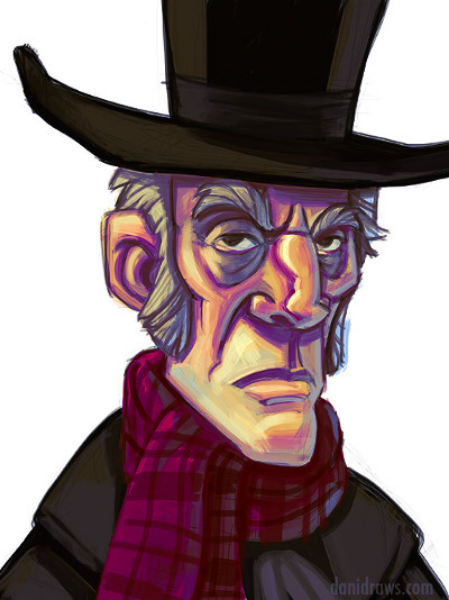}\\
 \centering
 \includegraphics[width=.3\linewidth]{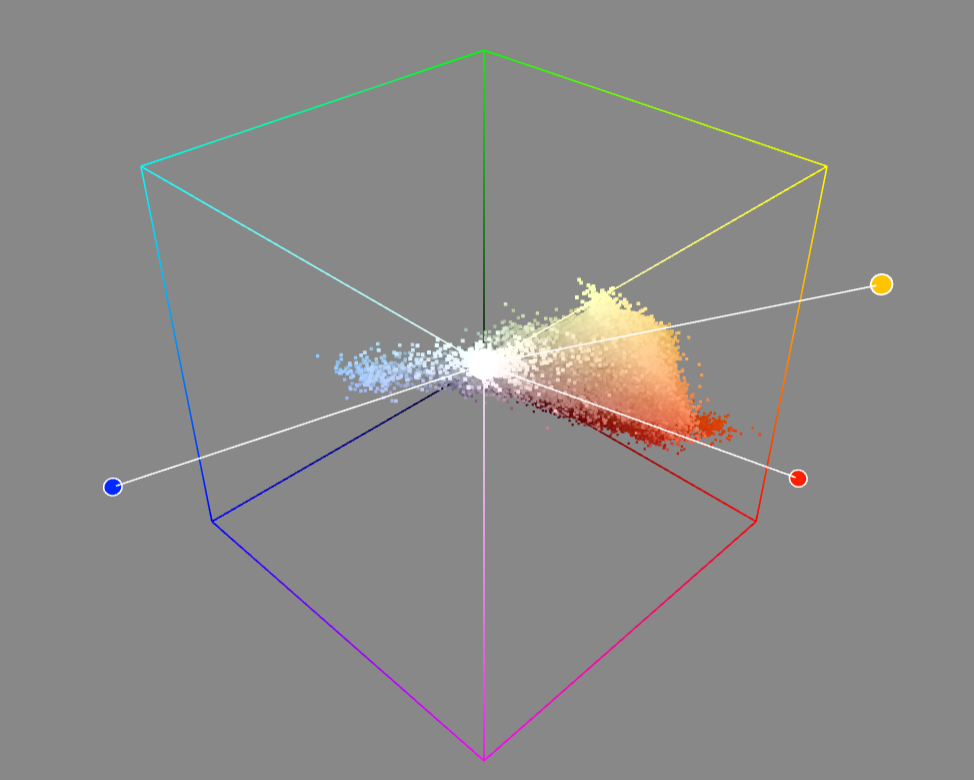}
 \includegraphics[width=.3\linewidth]{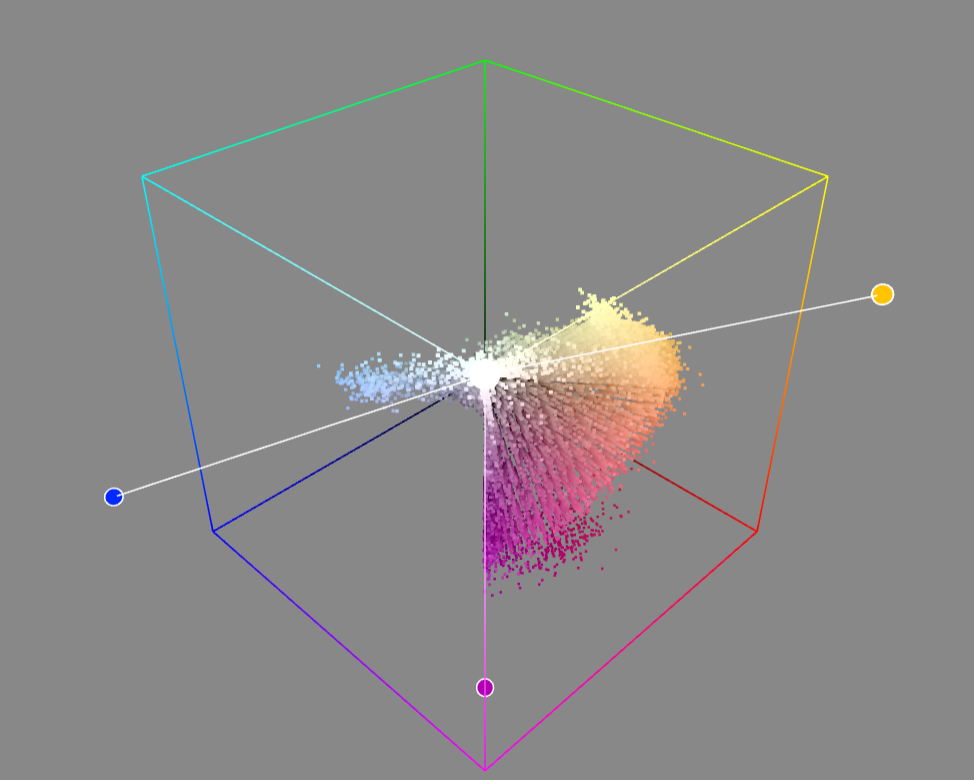}
 \includegraphics[width=.3\linewidth]{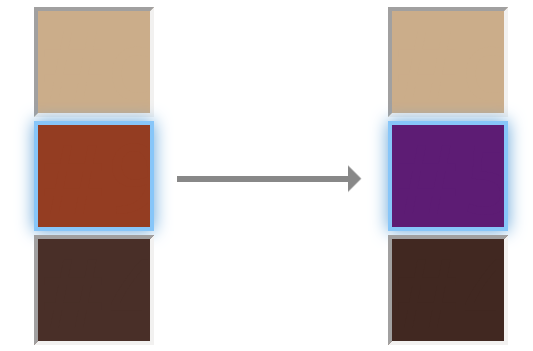}
\caption{ \label{fig:scrooge}
  Left: input drawing.
  Middle: recoloring using our approach, where the hue of the orange triangle is set to purple. Right: recoloring using color-palettes~\cite{Chang:2015:PPR}.
  }
\end{figure}

\textbf{Changing triangles color vertices}:
when rotating the triangle color vertices around the illuminant axis, color hues are modified. Thanks to the linear interpolation in cylindrical coordinates of the transformations, the color distribution presents no discontinuity and spread smoothly between the structures triangles (see Figure~\ref{fig:teaser}).
Our approach produces results comparable to the state of the art palette-based photo recoloring~\cite{Chang:2015:PPR}, as shown in Figure~\ref{fig:scrooge}.

\begin{figure}[t]
\centering%
\includegraphics[width=.49\linewidth]{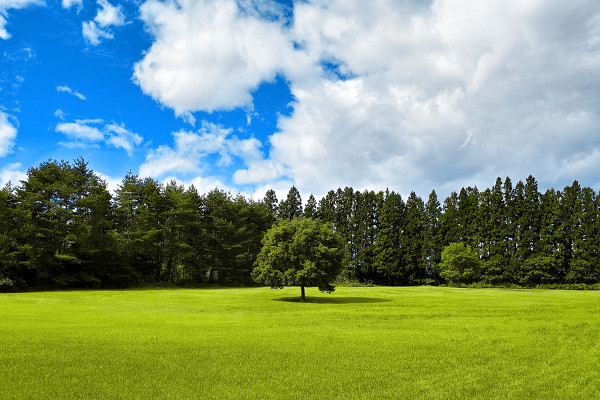}
\includegraphics[width=.49\linewidth]{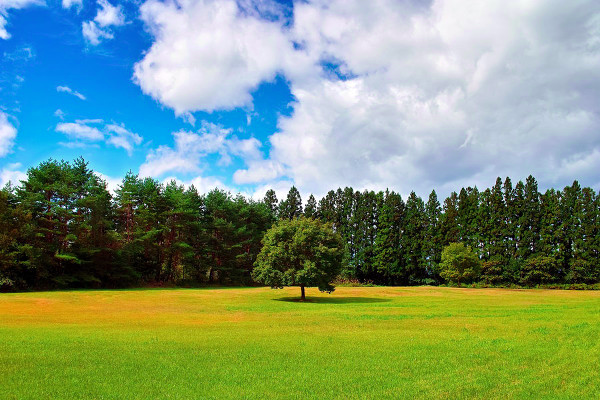}\\
\centering
\includegraphics[width=.49\linewidth]{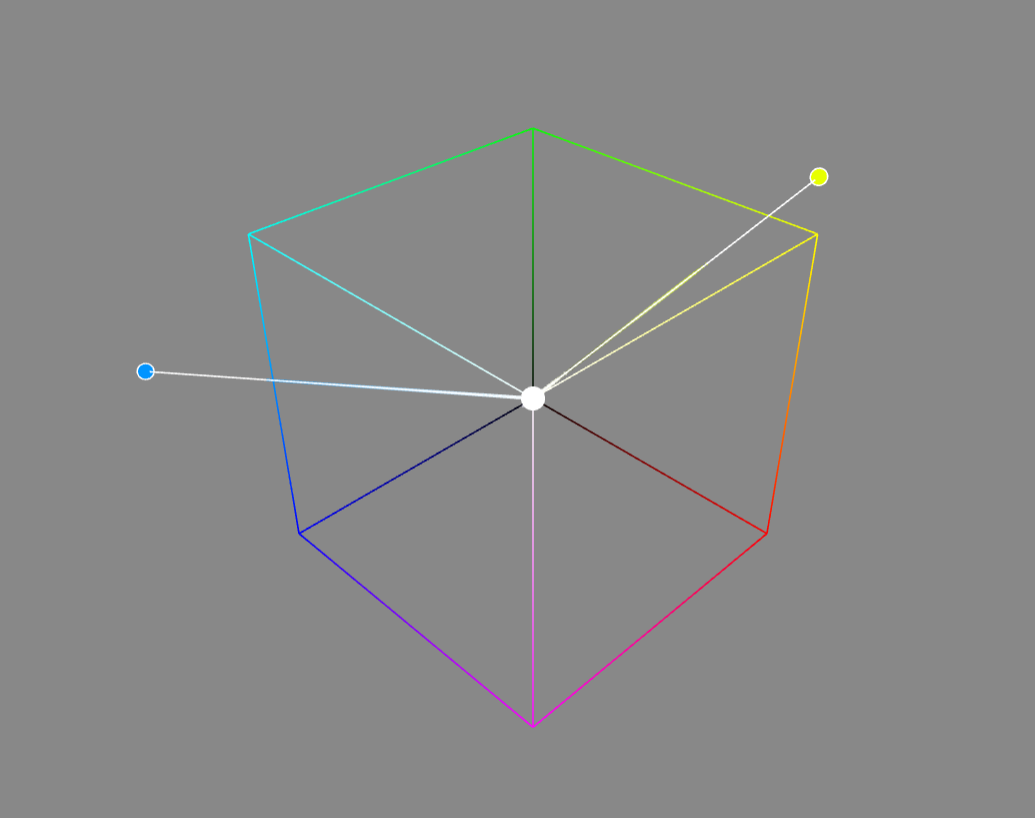}
\includegraphics[width=.49\linewidth]{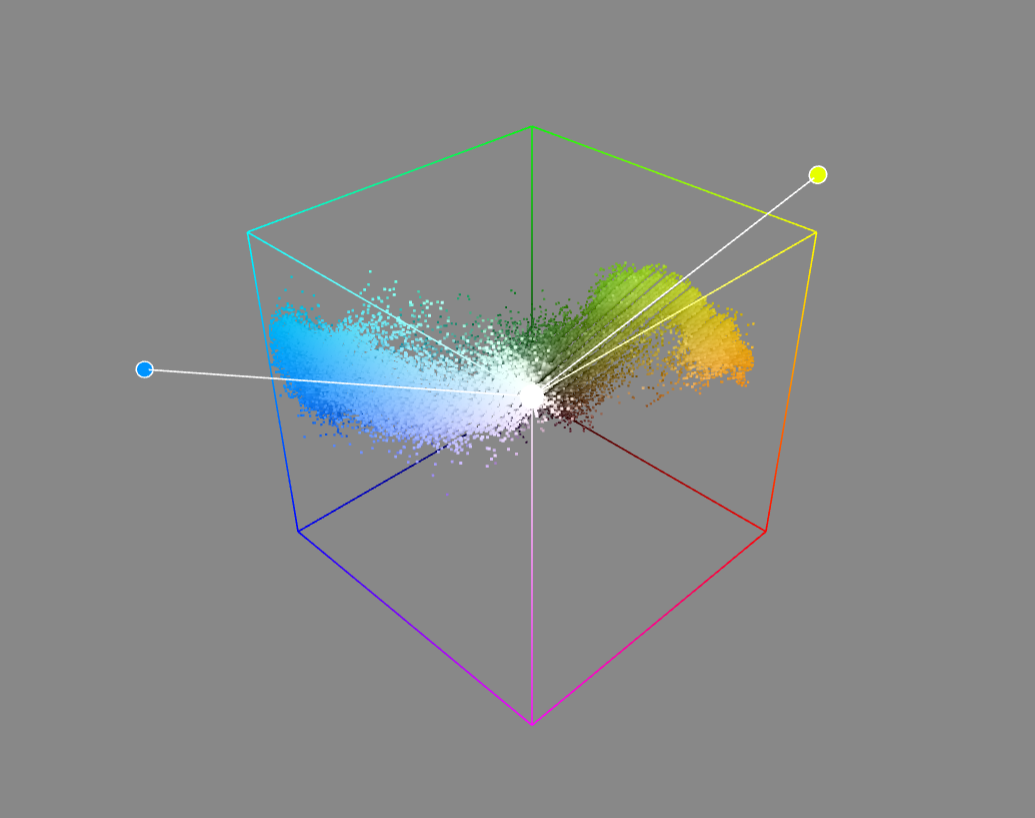}
\caption{
\label{fig:offset} See Figure~\ref{fig:colorpointscloud} for input image. Left: reduction of contrast (project on triangles). Right: boost contrasts (multiply point-triangle distances by 1.5).}
\end{figure}

\textbf{Structural filtering}:
Scaling the distance between color points and their nearest structure triangle controls the hue variations and color complexity in the image.
Projecting the colors on their nearest triangle lessen hue variation and removes hue diversity in textures.
On the other hand, when pushing the color distribution away from the triangle, the resulting colors span a larger area in color space and provides more vivid images (Figure~\ref{fig:offset}).


\section{Limitation and Future Work}

\textbf{Automatic extraction}:
in our current implementation, the extraction of the structure is limited to images with a single illuminant color, and require user inputs.
We believe that an automatic process could discover automatically the structure from the input point cloud, with a limited number of user inputs.

\noindent\textbf{Modelling other geometrical configurations and transformations}:
 images involving complex lighting interactions tends to generate point clouds that cannot be completely explained using triangular structures, e.g. the color bleeding in Figure~\ref{fig:cg}.
An interesting future work direction would be to explore other structural element shapes such that 3D parametric lines and surfaces.
In addition, we would also like to study other geometrical transformation,  
for instance to model material properties.


\noindent\textbf{Image layering}:
 we plan to investigate how to define image layers from our structure, for instance to weight the influence of point operators and other filters in imaging software.


\section{Conclusion}

We present a new geometric structure for color point clouds, which models the distribution of the colors of an image according to the physical properties of light/material interaction.
Our structure allows users to edit the colors of an image for recoloring or filtering.
We present a simple yet efficient optimization to instantiate a structure on a color point cloud.


\bibliographystyle{unsrt}

\bibliography{main}

\begin{thebibliography}{1}

\bibitem{Szeliski:2010:CVA:chapter3}
Richard Szeliski.
\newblock {\em Computer Vision: Algorithms and Applications}.
\newblock Springer-Verlag, Berlin, Heidelberg, 1st edition, 2010.

\bibitem{Solomon:2010:FDI}
Chris Solomon and Toby Breckon.
\newblock {\em Fundamentals of Digital Image Processing: A Practical Approach
  with Examples in Matlab}.
\newblock Wiley, 2010.

\bibitem{Szeliski:2010:CVA}
Richard Szeliski.
\newblock {\em Computer Vision: Algorithms and Applications}.
\newblock Springer-Verlag, Berlin, Heidelberg, 1st edition, 2010.

\bibitem{Tan:2016}
Jianchao Tan, Jyh-Ming Lien, and Yotam Gingold.
\newblock Decomposing images into layers via rgb-space geometry.
\newblock {\em ACM Trans. Graph.}, 36(1):7:1--7:14, November 2016.

\bibitem{Duchene:2017:MIA}
Sylvain Duch\^{e}ne, Carlos Aliaga, Tania Pouli, and Patrick P{\'e}rez.
\newblock Mixed illumination analysis in single image for interactive color
  grading.
\newblock In {\em Proceedings of the Symposium on Non-Photorealistic Animation
  and Rendering}, NPAR '17, pages 10:1--10:10, New York, NY, USA, 2017. ACM.

\bibitem{Chang:2015:PPR}
Huiwen Chang, Ohad Fried, Yiming Liu, Stephen DiVerdi, and Adam Finkelstein.
\newblock Palette-based photo recoloring.
\newblock {\em ACM Trans. Graph.}, 34(4):139:1--139:11, July 2015.

\bibitem{Mellado:2017:CPE}
Nicolas Mellado, David Vanderhaeghe, Charlotte Hoarau, Sidonie Christophe,
  Mathieu Br{\'e}dif, and Loic Barthe.
\newblock Constrained palette-space exploration.
\newblock {\em ACM Trans. Graph.}, 36(4):60:1--60:14, July 2017.

\bibitem{Tan:2018:EPD}
Jianchao Tan, Jose Echevarria, and Yotam Gingold.
\newblock Efficient palette-based decomposition and recoloring of images via
  rgbxy-space geometry.
\newblock {\em ACM Transactions on Graphics (TOG)}, 37(6):262:1--262:10,
  November 2018.

\bibitem{Kajiya:1986:RE}
James~T. Kajiya.
\newblock The rendering equation.
\newblock {\em SIGGRAPH Comput. Graph.}, 20(4):143--150, August 1986.

\end{thebibliography}

\newpage

\end{document}